\documentclass[pdflatex,sn-aps]{sn-jnl}

\usepackage{graphicx}%
\usepackage{multirow}%
\usepackage{amsmath,amssymb,amsfonts}%
\usepackage{amsthm}%
\usepackage{mathrsfs}%
\usepackage[title]{appendix}%
\usepackage{xcolor}%
\usepackage{textcomp}%
\usepackage{manyfoot}%
\usepackage{booktabs}%
\usepackage{algorithm}%
\usepackage{algorithmicx}%
\usepackage{algpseudocode}%
\usepackage{listings}%
\usepackage{newtx}
\usepackage{physics2}
\usephysicsmodule{ab, ab.braket}
\usepackage{bm}
\usepackage{ulem}
\usepackage{xcolor}
\usepackage{comment}
\DeclareRobustCommand{\erase}{\bgroup\markoverwith{\textcolor{red}{\rule[.5ex]{2pt}{0.4pt}}}\ULon}
\newcommand{\red}{\textcolor{black}}
\newcommand{\blue}{\textcolor{black}}

\newcommand{\so}{\Delta_{\mathrm{so}}}
\newcommand{\AM}{\Delta_{\mathrm{AM}}}
\newcommand{\kF}{k_{\mathrm{F}}}
\newcommand{\bmkF}{\bm{k}_{\mathrm{F}}}
\newcommand{\Tc}{T_{\mathrm{c}}}
\newcommand{\sgn}[1]{\mathrm{sgn}(#1)}
\newcommand{\fd}{f^{\dagger}}
\newcommand{\pert}[3]{#1_{#2,\lambda}^{(#3)}}
\newcommand{\fdpert}[3]{#1_{#2,\lambda}^{\dagger(#3)}}
\newcommand{\qst}{\bm{q}_{\mathrm{st}}}
\newcommand{\qdip}{\bm{q}_{\mathrm{dip}}}
\newcommand{\ratio}{|\Delta_{2\bm{Q}-\qst}/\Delta_{\qst}|}
\newcommand{\Qheli}{\bm{Q}}

\theoremstyle{thmstyleone}%
\theoremstyle{thmstyletwo}%

\theoremstyle{thmstylethree}%

\begin{document}
\title{Instability toward Superconducting Stripe Phase in Altermagnets with Strong Rashba Spin-Orbit Coupling}


\author[1]{\fnm{Kohei} \sur{Mukasa}}\email{mukasa.kohei.p7@dc.tohoku.ac.jp}

\author[1,2]{\fnm{Yusuke} \sur{Masaki}}\email{yusuke.masaki.c1@tohoku.ac.jp}

\affil[1]{\orgdiv{Department of Applied Physics}, \orgname{Tohoku University}, \orgaddress{\city{Sendai}, \postcode{980-8579}, \country{Japan}}}

\affil[2]{\orgdiv{Research and Education Center for Natural Science}, \orgname{Keio University}, \orgaddress{\street{Hiyoshi 4-4-1}, \city{Yokohama}, \postcode{223-8521}, \country{Japan}}}



\abstract{We numerically investigate finite-momentum superconductivity in noncentrosymmetric metallic altermagnets with $d$-wave spin-splitting and strong Rashba-type spin-orbit coupling. Focusing on a stripe phase in which Cooper pairs acquire multiple center-of-mass momenta, we construct phase diagrams that reveal phase boundaries between the stripe phase and a helical phase characterized by a single center-of-mass momentum. Our results show that the stripe phase emerges at low temperatures and exhibits a reentrant behavior as a function of the strength of the altermagnetic splitting. We further analyze the stripe phase within a linearized gap equation, and uncover the mechanism of the pairing formation unique to the stripe phase. This mechanism originates from the anisotropic deformation of the Fermi surfaces induced by the altermagnetic splitting, highlighting the intriguing interplay between the spin-orbit coupling and the altermagnets.}




\maketitle
\section{Introduction}\label{introduction}
\begin{figure}[tb]
    \centering
    \includegraphics[width=\linewidth]{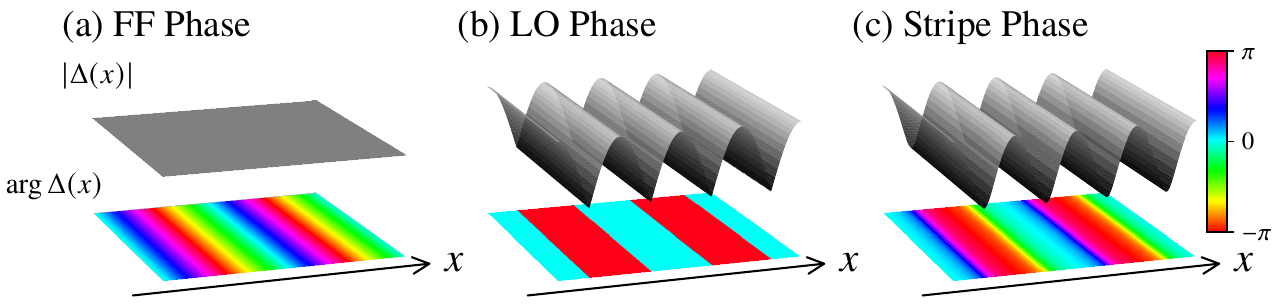}
    \caption{\red{Schematic form of the spatially modulated superconducting states. (a) FF phase, (b) LO phase and (c) stripe phase with one-dimensional modulation. Gray scale surface plots indicate the amplitude of the order parameter $|\Delta(x)|$ and the color maps indicate the argument of $\Delta(x)$.}}
    \label{fig:schematic_of_Delta}
\end{figure}

The conventional superconductors described by the BCS theory energetically favor the zero center-of-mass momentum of Cooper pairs due to the spin degeneracy of the Fermi surfaces (FSs). 
However, some spin-splitting effects such as uniform magnetic fields can lead to the Cooper pairs with finite momenta.
Such superconducting states are known as the finite-momentum superconductivity and lots of studies have been working on them.
The finite-momentum superconductivity with a momentum $\bm{q}$ has a modulation of the superconducting order parameter $\Delta(\bm{R})$ in real space \red{with a center-of-mass coordinate $\bm{R}$}.
While the FF state proposed by Fulde and Ferrell~\cite{FFstate} is described by a phase-modulated order-parameter $\Delta(\bm{R})=\Delta_{\bm{q}}e^{i\bm{q}\cdot\bm{R}}$[Fig.~\ref{fig:schematic_of_Delta}(a)], the LO state proposed by Larkin and Ovchinnikov~\cite{LOstate} is described by an amplitude-modulated order-parameter $\Delta(\bm{R})=2\Delta_{\bm{q}}\cos{\bm{q}\cdot\bm{R}}$[Fig.~\ref{fig:schematic_of_Delta}(b)].
Generally, in low temperature and high magnetic field region, the LO state is more stable than the FF state because the amplitude modulation can lower the increase in energy due to paramagnetic effects.
\par
In noncentrosymmetric superconductors, the FSs can be shifted by the coupling of the antisymmetric spin-orbit coupling and the in-plane magnetic field, leading to the finite-momentum superconductivity known as the helical phase.
The helical phase also has a phase modulation in the phase part as the FF state does, and it has attracted considerable attention in connection to the intrinsic mechanism of the superconducting diode effect, which is a nonreciprocal phenomenon of supercurrents~\cite{Daido_Ikeda_Yanase_2022,Yuan_Fu_2022,He_Tanaka_Nagaosa_2022, Ilić_Bergeret_2022}.
\red{Moreover, the superconducting states described by $\Delta(\bm{R})=\Delta_{\bm{q}}e^{i\bm{q}\cdot\bm{R}}+\Delta_{-\bm{q}}e^{-i\bm{q}\cdot\bm{R}}+\cdots$ can be stabilized even in noncentrosymmetric systems~\cite{Barzykin_2002, Agterberg-helicalstripe-2007,Dimitrova_Feigel’man_2007, Agterberg_Babaev_Garaud_2014, Aoyama_2024, Xie_Chai_Zhu_Zha_2024} as the LO phase is under uniform magnetic fields.
Such a state is known as the stripe phase and it involves the spatial modulation of both the phase and the amplitude of $\Delta(\bm{R})$ unlike the FF state and the LO state.}
However, the study \red{of the stripe phase} in noncentrosymmetric superconductors is limited.
\par
Altermagnets are a newly discovered class of magnets which host an anisotropic spin-splitting in electronic energy bands, and numerous studies have been devoted to enhance functionality of the altermagnets\cite{Ahn_Hariki_Lee_Kuneš_2019, Naka_Hayami_Kusunose_Yanagi_Motome_Seo_2019, Hayami_Yanagi_Kusunose_2019, Naka_Hayami_Kusunose_Yanagi_Motome_Seo_2020, Naka_Motome_Seo_2021, Šmejkal_Sinova_Jungwirth_2022}.
Interestingly, altermagnetic metals have been theoretically shown to host the finite-momentum superconductivity due to their spin-splitting FSs\cite{Soto-Garrido_Fradkin_2014,Sumita_Naka_Seo_2023, Banerjee_Scheurer_2024,Chakraborty_Black-Schaffer_2024,Hong_Park_Kim_2025,Mukasa_Masaki_2025,Sim_Knolle_2025,Iorsh_2025, Liu_Hu_Liu_2025, Sumita_Naka_Seo_2025}.
As the helical phase is stabilized by the coupling of the magnetic field and the RSOC, Ref.~\cite{Mukasa_Masaki_2025} also revealed that the coupling of the altermagnetic splitting and the RSOC can induce the finite-momentum superconductivity with a single-momentum.
However, the possibility of the stripe phase in the presence of the altermagnetic splitting still remains to be investigated.
\par
In this work, we address the stripe phase in the superconducting altermagnets with the strong RSOC using a quasiclassical framework.
We numerically find that the stripe phase displays a reentrant behavior as a function of the strength of the altermagnetic spin-splitting, which is in stark contrast to the previous studies on the Rashba superconductors in the presence of Zeeman magnetic fields (hereafter referred to as the Rashba--Zeeman superconductors)~\cite{Dimitrova_Feigel’man_2007, Agterberg-helicalstripe-2007, Agterberg_Babaev_Garaud_2014, Aoyama_2024}.
We further investigate properties of the stripe phase such as involving momenta and a superconducting gap structure within a linearized gap equation, and reveal that there are distinct mechanisms for the stripe phase depending on the strength of the altermagnetic splitting.
By combining numerical analyses with the physical geometry of the FSs shaped by the altermagnetic spin splitting and the RSOC, our study provides insight into the emergence of the finite-momentum superconductivity involving multiple center-of-mass momentum in altermagnets.
\section{Model and Method}\label{modelandmethod}
We consider two-dimensional $d$-wave altermagnets with the RSOC and the spin-singlet superconductivity.
The microscopic Hamiltonian is
\begin{gather}
    \hat{H} = \sum_{\bm{k}}\sum_{\sigma\sigma'}\hat{c}^{\dagger}_{\bm{k}\sigma}[H_0(\bm{k})]_{\sigma\sigma'}\hat{c}_{\bm{k}\sigma'}
    +\sum_{\bm{k},\bm{k}',\bm{q}}V(\bm{k},\bm{k}')\hat{c}^{\dagger}_{\bm{k}+\bm{q}/2\uparrow}\hat{c}^{\dagger}_{-\bm{k}+\bm{q}/2\downarrow}\hat{c}_{-\bm{k}'+\bm{q}/2\downarrow}\hat{c}_{\bm{k}'+\bm{q}/2\uparrow},\\
    \label{eq:Normal_H0}
    H_0(\bm{k}) = \xi_k + \ab[\so\bm{e}_z\times\bar{\bm{k}} + \AM(\bar{k}^2_x-\bar{k}^2_y)\bm{n}]\cdot\bm{\sigma},
\end{gather}
where $\hat{c}_{\bm{k}\sigma}$($\hat{c}^{\dagger}_{\bm{k}\sigma}$) is the annihilation (creation) operator of an electron with momentum $\bm{k}$ and spin $\sigma =\uparrow, \downarrow$, and $V(\bm{k},\bm{k}')$ is an attractive interaction between two electrons.
The normal-state Hamiltonian is described by $H_0(\bm{k})$, which includes the spin-independent kinetic energy term $\xi_k=k^2/2m - \mu$, the RSOC term with strength $\so$, and the altermagnetic splitting term with strength $\AM$ and the N\'{e}el vector $\bm{n}$.
Here, $\mu$ is the chemical potential, $m$ is the mass of an electron, $\bar{\bm{k}}=\bm{k}/\kF$ with $\kF=\sqrt{2\mu m}$, and $\bm{\sigma}=(\sigma_x,\sigma_y,\sigma_z)$ denotes the Pauli matrices.
We introduce $\bm{g}(\bm{k})= \bm{e}_z\times\bar{\bm{k}}$ and $\bm{J}(\bm{k}) = \AM(\bar{k}^2_x-\bar{k}^2_y)\bm{n}$ for later convenience.
We set $\hbar=k_{\mathrm{B}}=1$ throughout this work.\par
Here we outline the assumptions in this work.
First, we do not consider sublattice degrees of freedom which is important in altermagnetic materials. 
Instead, we phenomenologically incorporate the $d$-wave altermagnetic spin splitting in a present single-band model and treat the effect of the altermagnet on FSs within the quasiclassical theory.
\red{Such simplification is valid when we focus on the system having the FSs only around the $\Gamma$ point in its momentum space, which can be realized in the minimal model of the altermagnet~\cite{Brekke_Brataas_Sudbø_2023}.
In the case that the FSs are far away from the $\Gamma$ point, the continuum representation of the altermagnetic splitting in Eq.~\eqref{eq:Normal_H0} is no longer justified.
Indeed, Ref.~\cite{Sumita_Naka_Seo_2025} pointed out that the difference in the momentum dependence of the altermagnetic splitting can qualitatively change the properties of the finite-momentum superconductivity.}
Second, \red{we adopt self-consistent calculation only for the superconducting order parameter and incorporate the altermagnetic order as an external parameter.
We note that Ref.~\cite{Chakraborty_Black-Schaffer_2025_constraints_on} theoretically found the coexistence of the altermagnetic order and the finite-momentum superconductivity by solving self-consistent equations for both orders in a minimal model of the altermagnet.}
\red{As for the energy scale, }we assume $\AM \ll \so \ll \mu$ and neglect an inter-band pairing for simplicity, although experimental situations which meet this energy scale are thought to be rare. 
Comprehensive calculations including both the inter-band pairing and the intra-band pairing are future problems.
Third, we focus only on the case in which the N\'{e}el vector of the altermagnet is in the plane. Hence, the following calculations, we set $\bm{n}=\bm{e}_y$.
As for the symmetry of the superconductivity, we focus on the $d_{x^2-y^2}$-wave superconducting order parameter, where its node directions are the same as those of the altermagnetic splitting.
The $d_{x^2-y^2}$-wave finite-momentum superconductivity in the altermagnets is theoretically reported\cite{Chakraborty_Black-Schaffer_2024,Mukasa_Masaki_2025}.
\par
Because of the second assumption, it is convenient to adopt the basis which diagonalizes the RSOC term, referred to as the RSOC basis in this work. For this purpose, we employ the same quasiclassical framework introduced in Ref. \cite{Agterberg-helicalstripe-2007} for the Rashba--Zeeman superconductors, and explore the finite-momentum superconductivity with multi-$\bm{q}$ in the present model. 
The Eilenberger equations in the RSOC basis are given as follows:
\begin{align}
\label{eq:Eilenberger_f}
    \ab[\omega_n +i\lambda\hat{\bm{g}}(\bmkF)\cdot\bm{J}(\bmkF)+\frac{1}{2}\bm{\varv}(\bmkF)\cdot \bm{\nabla}]f_{\lambda}(\omega_n, \bmkF, \bm{R}) &= \Delta(\bmkF,\bm{R})g_{\lambda}(\omega_n, \bmkF, \bm{R}), \\
\label{eq:Eilenberger_fdg}
    \ab[\omega_n +i\lambda\hat{\bm{g}}(\bmkF)\cdot\bm{J}(\bmkF)-\frac{1}{2}\bm{\varv}(\bmkF)\cdot \bm{\nabla}]f^{\dagger}_{\lambda}(\omega_n, \bmkF, \bm{R}) &= \Delta^*(\bmkF,\bm{R})g_{\lambda}(\omega_n, \bmkF, \bm{R}),
\end{align}
with the quasiclassical Green's functions $g_{\lambda}(\omega_n, \bmkF, \bm{R})$, $f_{\lambda}(\omega_n, \bmkF, \bm{R})$, and $f_{\lambda}^{\dagger}(\omega_n, \bmkF, \bm{R})$ satisfying the normalization condition 
\begin{equation}
    \label{eq:normalization}
    \ab[g_{\lambda}(\omega_n, \bmkF, \bm{R})]^2+f_{\lambda}(\omega_n, \bmkF, \bm{R})f^{\dagger}_{\lambda}(\omega_n, \bmkF, \bm{R})=1.
\end{equation}
In our notation, $\lambda = + (-)$ represents the inner (outer) FS. In the derivation of the equations, the basis-mixing term originating from the $\bm{J}(\bm{k})\cdot\bm{\sigma}$ has been neglected, which is equivalent to ignoring the inter-band pairing.
Here, $\omega_n=(2n+1)\pi T$ are the fermionic Matsubara frequencies, $\bmkF$ is the Fermi momentum, $\bm{R}$ is the center-of-mass coordinate of Cooper pairs, and $\hat{\bm{g}}(\bm{k})$ is the normalized $\bm{g}(\bm{k})$.
The following symmetry relations in terms of $\omega_n$ hold: $g_{\lambda}(\omega_n)=-[g_{\lambda}(-\omega_n)]^*$ and $f_{\lambda}(\omega_n) = [f^{\dagger}_{\lambda}(-\omega_n)]^*$.
We assume the separable form $V(\bmkF, \bmkF')=-V\psi_{\Gamma}(\bmkF)\psi^*_{\Gamma}(\bmkF')$ ($V  > 0$) and $\Delta(\bmkF, \bm{R})=\Delta(\bm{R})\psi_{\Gamma}(\bmkF)$ using the form factor $\psi_{\Gamma}(\bm{k})=\sqrt{2}\cos{2\phi_{\bm{k}}}$ for the $d_{x^2-y^2}$-wave superconductivity with the azimuthal angle $\phi_{\bm{k}}$, and then we get the superconducting gap equation within the mean-field approximation given as
\begin{equation}
\label{eq:gap_eq_real_space}
    \Delta(\bm{R}) 
    =\frac{\pi T V}{2} \sum_{n,\lambda}N_\lambda\braket<\psi^*_{\Gamma}(\bmkF)f_{\lambda}(\bmkF,\bm{R},\omega_n)>_{\mathrm{FS}},
\end{equation}
where $\ab<\cdots>_{\mathrm{FS}}$ denotes the average on the FS defined as
\begin{equation}
    \ab<h(\bmkF)>_{\mathrm{FS}} = \int^{2\pi}_0 \frac{d\phi_{\bm{k}}}{2\pi}h\ab(k_{\mathrm{F}}\cos\phi_{\bm{k}},k_{\mathrm{F}}\sin\phi_{\bm{k}}).
\end{equation}
Here, we have introduced the band-dependent density of states at the Fermi level as
\begin{equation}
    N_{\lambda} = N_0(1-\lambda \delta N),
\end{equation}
where $N_0$ is the average density of states and $\delta N= \so/2\mu$ describes the difference in the density of states between two bands $\lambda=\pm$ [$\delta N = (N_- - N_+)/2N_0$].
\red{Note that $\AM$ and $\so$ have a clearly separated  energy hierarchy when $\delta N\ne0$, since $\AM/\so = (\AM/\Tc)/(\so/\mu)\times (\Tc/\mu) \to 0$ in the quasiclassical limit $\Tc/\mu \to 0$.}
Hereafter, we measure the strength of the RSOC via $\delta N$ as introduced in Refs.~\cite{Agterberg-helicalstripe-2007,Agterberg_Babaev_Garaud_2014,Aoyama_2024,Ilić_Bergeret_2022}.
The free energy functional is given as 
\begin{gather}
    \label{eq:F_functional}F_{\mathrm{SN}}=\frac{1}{\Omega}\int d\bm{R} \ab [\frac{2|\Delta(\bm{R})|^2}{V}-\pi T \sum_{n,\lambda}N_{\lambda}\ab<I_{\lambda}(\bmkF,\bm{R},\omega_n) >_{\mathrm{FS}}], \\
    \label{eq:I_def}I_{\lambda}(\bmkF,\bm{R},\omega_n) = \frac{\Delta^*(\bmkF,\bm{R})f_{\lambda}(\bmkF,\bm{R},\omega_n)+\fd_{\lambda}(\bmkF,\bm{R},\omega_n)\Delta(\bmkF,\bm{R})}{1+\sgn{\omega_n}g_{\lambda}(\bmkF,\bm{R},\omega_n)},
\end{gather}
where $\Omega$ is the volume of the system.
\par
We solve Eqs. \eqref{eq:Eilenberger_f} and \eqref{eq:Eilenberger_fdg} by the Fourier transform in terms of the center-of-mass momentum $\bm{q}$:
\begin{align}
    \Delta(\bmkF, \bm{R}) &= \sum_{\bm{p}} \Delta_{\bm{p}}(\bmkF)e^{i\bm{p}\cdot\bm{R}}, \\
    x^{}_{\lambda}(\omega_n, \bmkF, \bm{R}) &= \sum_{\bm{p}} x^{}_{\bm{p},\lambda}(\omega_n, \bmkF)e^{i\bm{p}\cdot\bm{R}}, 
\end{align}
where $x$ denotes $g$, $f$, and $f^{\dagger}$.
Note that these Fourier components also follow the symmetry $g_{\bm{p},\lambda}(\omega_n)=-[g_{-\bm{p},\lambda}(-\omega_n)]^*$ and $f_{\bm{p},\lambda}(\omega_n)=[f^{\dagger}_{-\bm{p},\lambda}(-\omega_n)]^*$.
Substituting the above expansions into Eqs. \eqref{eq:Eilenberger_f}--\eqref{eq:gap_eq_real_space}, we obtain
\begin{gather}
    \label{eq:f_Eilenberger_fourier}
    \omega_{\bm{q}',\lambda} f_{\bm{q}',\lambda} = \sum_{\bm{p}}\Delta_{\bm{p}}(\bmkF)g_{\bm{q}'-\bm{p},\lambda}, \\
    \label{eq:fdg_Eilenberger_fourier}
    \omega_{\bm{q}',\lambda} f^{\dagger}_{-\bm{q}',\lambda} = \sum_{\bm{p}}\Delta^*_{\bm{p}}(\bmkF)g_{-\bm{q}'+\bm{p},\lambda}, \\
    \label{eq:normalize_fourier}\sum_{\bm{p}} \ab [g_{\bm{p}+\bm{q}',\lambda}g_{-\bm{p},\lambda}+f_{\bm{p}+\bm{q}',\lambda}f^{\dagger}_{-\bm{p},\lambda}]=\delta_{\bm{q}',\bm{0}},\\
    \label{eq:gap_eq_momentum_space}
    \Delta_{\bm{p}}(\bmkF) = \dfrac{\pi T V}{2}\sum_{n,\lambda}
    N_{\lambda}\braket<\psi_{\Gamma}^*(\bmkF)f_{\bm{p},\lambda}(\omega_n,\bmkF)>_{\mathrm{FS}}, 
\end{gather}
with $\omega_{\bm{p}, \lambda}=\omega_n +i\lambda\hat{\bm{g}}(\bmkF)\cdot\bm{J}(\bmkF)+i\bm{\varv}(\bmkF)\cdot\bm{p}/2$.\par
Let us expand the superconducting order parameter as $\Delta(\bmkF, \bm{R})=\Delta^{(0)}_{\bm{Q}}(\bmkF)e^{i\bm{Q}\cdot\bm{R}}+\Delta^{(1)}_{\bm{q}}(\bmkF)e^{i\bm{q}\cdot\bm{R}}+\Delta^{(1)}_{2\bm{Q}-\bm{q}}(\bmkF)e^{i(2\bm{Q}-\bm{q})\cdot\bm{R}}$. The first term denotes the helical superconductivity minimizing the free energy within the single-$\bm{q}$ approximation and the latter two perturbative terms represent the instability toward the stripe phase.
We also expand $g_{\bm{p},\lambda}$, $f_{\bm{p},\lambda}$ and $f^{\dagger}_{\bm{p},\lambda}$ as $x_{\bm{p},\lambda} = \pert{x}{\bm{p}}{0}\delta_{\bm{p},\bm{Q}} + \pert{x}{\bm{p}}{1} + \pert{x}{\bm{p}}{2} + \cdots$ with $x_{\bm{p},\lambda} = g_{\bm{p}-\bm{Q},\lambda}$, $f_{\bm{p},\lambda}$, $f_{-\bm{p},\lambda}^{\dagger}$ to perturbatively solve Eqs.~\eqref{eq:f_Eilenberger_fourier} -- \eqref{eq:normalize_fourier} in terms of $\Delta_{\bm{q}}^{(1)}$ and $\Delta_{2\bm{Q}-\bm{q}}^{(1)}$.
We first calculate the helical phase having the superconducting gap $\Delta_{\bm{Q}}^{(0)}$ with a single momentum $\bm{Q}$, which minimizes the free energy \eqref{eq:F_functional}. The quasiclassical Green's functions for the helical phase are given by
\begin{align}
    \pert{g}{\bm{0}}{0} &= \frac{\omega_{\bm{Q},\lambda}}{\sqrt{\omega_{\bm{Q},\lambda}^2+|\Delta^{(0)}_{\bm{Q}}(\bm{k}_\mathrm{F})|^2}},\quad 
    \label{eq:f_Q_0_ans}
    \pert{f}{\bm{Q}}{0} = \frac{\Delta^{(0)}_{\bm{Q}}(\bm{k}_{\mathrm{F}})}{\omega_{\bm{Q},\lambda}}\pert{g}{\bm{0}}{0},\quad\fdpert{f}{-\bm{Q}}{0} = \frac{\Delta^{*(0)}_{\bm{Q}}(\bm{k}_{\mathrm{F}})}{\omega_{\bm{Q},\lambda}}\pert{g}{\bm{0}}{0},
\end{align}
and \red{the gap equation~\eqref{eq:gap_eq_momentum_space} is reduced to}
\begin{align}
    \red{\label{eq:gapeq_in_helical}
    1 = \frac{\pi TV}{2}\sum_{n,\lambda}N_{\lambda}\ab<\frac{|\psi_{\Gamma}(\bmkF)|^2}{\sqrt{\omega_{\bm{Q},\lambda}^2+|\Delta^{(0)}_{\bm{Q}}(\bm{k}_\mathrm{F})|^2}}>_{\mathrm{FS}}
    \equiv \sum_{\lambda}\braket<K_{\lambda}(\bmkF)>_{\mathrm{FS}}.}
\end{align}
Then, we set the helical phase as the non-perturbative state, and perturbatively analyze the instability toward the stripe phase holding two additional modes $\Delta^{(1)}_{\bm{q}}$ and $\Delta^{(1)}_{2\bm{Q}-\bm{q}}$.
By substituting the above expansions into Eqs.~\eqref{eq:f_Eilenberger_fourier} -- \eqref{eq:normalize_fourier}, we obtain
\begin{equation}
    \label{eq:f_q_1st}
    \pert{f}{\bm{q}}{1} = \pert{g}{\bm{0}}{0}\frac{\ab(2\omega_{2\bm{Q}-\bm{q},\lambda}\omega_{\bm{Q},\lambda}+\ab|\Delta^{(0)}_{\bm{Q}}(\bmkF)|^2)\Delta^{(1)}_{\bm{q}}(\bmkF)-\ab\{\Delta^{(0)}_{\bm{Q}}(\bmkF)\}^2\ab(\Delta^{(1)}_{2\bm{Q}-\bm{q}}(\bmkF))^*}{2\omega_{\bm{q},\lambda}\omega_{2\bm{Q}-\bm{q},\lambda}\omega_{\bm{Q},\lambda}+\ab|\Delta^{(0)}_{\bm{Q}}(\bmkF)|^2\ab(\omega_{\bm{q},\lambda}+\omega_{2\bm{Q}-\bm{q},\lambda})},
\end{equation}
For simplicity, we denote the coefficients of $\Delta^{(1)}_{\bm{q}}$ and $(\Delta^{(1)}_{2\bm{Q}-\bm{q}})^*$ by $A_{\bm{q},\lambda}(\bmkF,i\omega_n)$ and $B_{\bm{q},\lambda}(\bmkF,i\omega_n)$, respectively, so that 
\begin{equation}
   \label{eq:AB_label}
   \pert{f}{\bm{q}}{1}
   = A_{\bm{q},\lambda}(\bmkF,i\omega_n)\Delta^{(1)}_{\bm{q}}
   + B_{\bm{q},\lambda}(\bmkF,i\omega_n)\ab(\Delta^{(1)}_{2\bm{Q}-\bm{q}})^*.
\end{equation}
We linearize the gap equation \eqref{eq:gap_eq_momentum_space} by using the above expression, and obtain
\begin{equation}
\label{eq:linearized_gap_eq}
    \delta\Vec{\Delta}^{(1)}_{\bm{q}} = \mathsf{M}(\bm{q})\delta\Vec{\Delta}^{(1)}_{\bm{q}},
\end{equation}
with $\delta\Vec{\Delta}^{(1)}_{\bm{q}}=\ab(\Delta_{\bm{q}}^{(1)}, (\Delta^{(1)}_{2\bm{Q}-\bm{q}})^*)^{\mathsf{T}}$.
The coefficient matrix $\mathsf{M}(\bm{q})$ is given by
\begin{equation}
\label{eq:def_M}
    \mathsf{M}(\bm{q})=\frac{\pi TV}{2}\sum_{n,\lambda}
    N_\lambda
    \begin{pmatrix}
        \ab<\psi^*_{\Gamma}(\bmkF)A_{\bm{q},\lambda}(\bmkF,i\omega_n)>_{\mathrm{FS}} & \ab<\psi^*_{\Gamma}(\bmkF)B_{\bm{q},\lambda}(\bmkF,i\omega_n)>_{\mathrm{FS}}\\
        \ab<\psi_{\Gamma}(\bmkF)B^*_{2\bm{Q}-\bm{q},\lambda}(\bmkF,i\omega_n)>_{\mathrm{FS}} & \ab<\psi_{\Gamma}(\bmkF)A^*_{2\bm{Q}-\bm{q},\lambda}(\bmkF,i\omega_n)>_{\mathrm{FS}}
    \end{pmatrix}.
\end{equation}
The smaller eigenvalue $\epsilon_1(\bm{q})$ of $\mathsf{I}_2 - \mathsf{M}(\bm{q})$ can be expressed as
\begin{equation}
\label{eq:small_eigenvalue}
    \epsilon_1(\bm{q}) = \frac{\ab(1-M_{11})+\ab(1-M_{22})}{2} - \sqrt{\ab(\frac{M_{11}-M_{22}}{2})^2 + \ab(M_{12})^2},
\end{equation}
where $\mathsf{I}_2$ is the $2\times 2$ identity matrix and $M_{ij} (i,j=1,2)$ are the elements of the matrix $\mathsf{M}$.
A negative $\epsilon_1(\bm{q})$ indicates the existence of the instability toward the stripe phase with the eigenstate $\delta\Vec{\Delta}^{(1)}_{\bm{q}}$ which lowers the free energy up to the second-order of $\Delta_{\bm{q}}^{(1)}$ and $(\Delta^{(1)}_{2\bm{Q}-\bm{q}})^*$.
Therefore, the phase boundary of the stripe phase with the additional momentum $\bm{q}_{\mathrm{st}}$ is determined by the condition $\epsilon_1(\bm{q}_{\mathrm{st}})=\min_{\bm{q}}\epsilon_1(\bm{q})=0$.
\par
Throughout our calculation, we assume the center-of-mass momentum in the $x$ direction\red{, and its $x$-component is written without bold font, e.g.,  $\bm{q}=q\bm{e}_x$.}
Energy, velocity, and momentum are, respectively, scaled by $\Tc$, $\varv_{\mathrm{F}}$, and $q_0$.
Here, $\Tc$ is the superconducting critical temperature without $\AM$, $\varv_{\mathrm{F}}$ is the Fermi velocity and $q_0 = \Delta_0/\varv_{\mathrm{F}}$ is the inverse of the coherence length with $\Delta_0/\Tc = \pi/e^{\gamma}$.
\red{We use $(N_0 V)^{-1}=\ln{\ab(T/\Tc)}+2\pi T\sum_{\omega_n>0}(\omega_n)^{-1}$ and set the cutoff energy $\omega_{\mathrm{c}}$ for the summation over Matsubara frequencies to $80\Tc$.}
\section{Results}\label{results}
\begin{figure}[t]
    \centering
    \includegraphics[width=\linewidth]{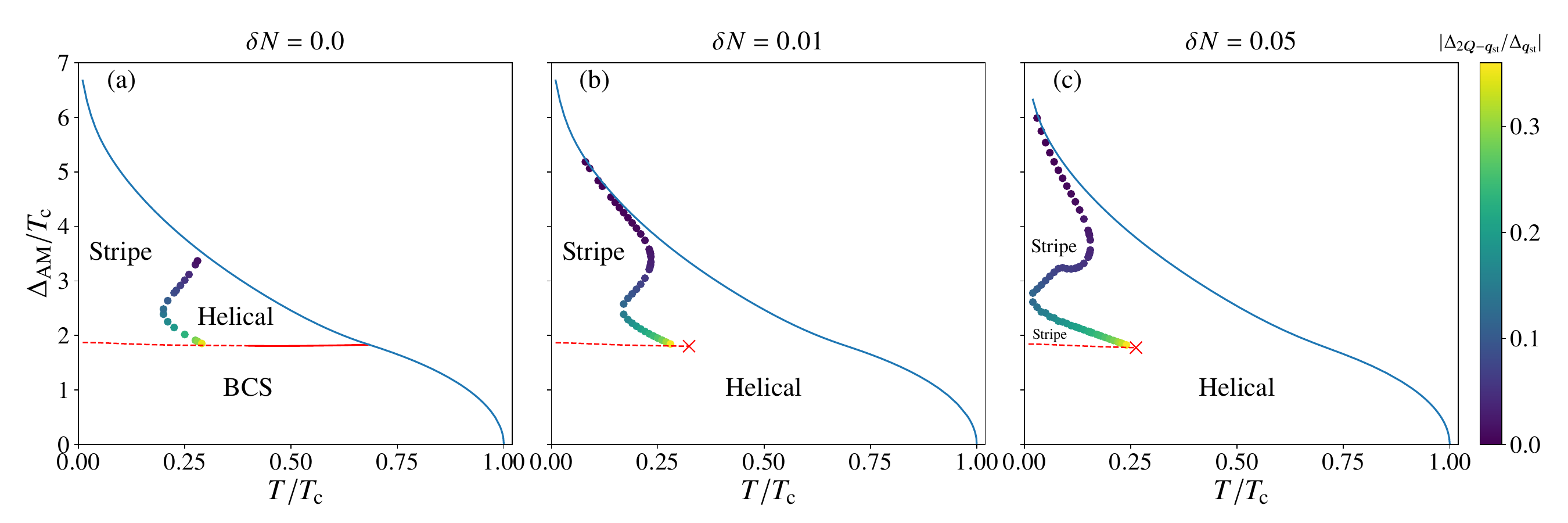}
    \caption{Phase diagrams in the $(T,\AM)$ plane for (a) $\delta N=0.0$, (b) $\delta N=0.01$, (c) $\delta N=0.05$. The blue lines show second-order transitions between superconducting and normal states. The dashed red lines indicate first-order transitions of $Q$; the red solid line in (a) shows its second-order transition. The red crosses in (b) and (c) mark the crossover onset. Scatter plots show the phase boundaries between the stripe phase and the helical phase. Their colors indicate the absolute value of $\ratio$ calculated from the gap equation Eq.~\eqref{eq:linearized_gap_eq}. \red{Note that the $\delta N=0$ case is not suitable for our framework because the inter-band pairing is not negligible as illustrated in the main text.}}
    \label{fig:phase_diagrams}
\end{figure}

\begin{figure}[tb]
    \centering
    \includegraphics[width=\linewidth]{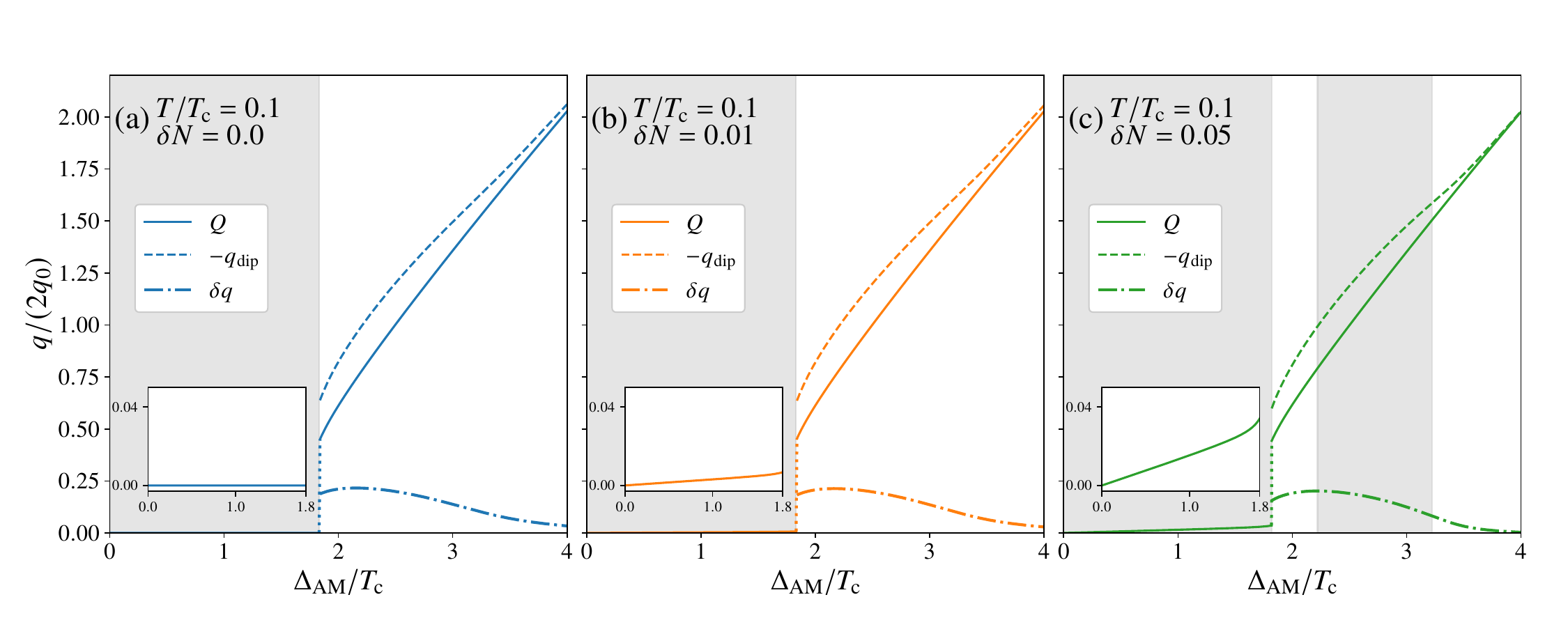}
    \caption{$\AM$ dependence of $Q$ and $q_{\mathrm{dip}}$ for three $\delta N$ values. The solid, dashed, and dash-dotted lines represent $Q$, $-q_{\mathrm{dip}}$, and $|\delta q|=||Q|-|q_{\mathrm{dip}}||$, respectively. The temperature is fixed at $T=0.1\Tc$. The dotted lines indicate \red{discontinuous changes} of $Q$. The inset shows an enlarged view in the region $0\le\AM/\Tc\le1.8$ for visibility. The shaded region indicates where the ground state is the helical phase.}
    \label{fig:AMdep_of_momentum}
\end{figure}

Our main results are summarized in Fig.~\ref{fig:phase_diagrams}, where the phase diagrams in the $(T,\AM)$ plane for three $\delta N$ values are presented.
Although the case of $\delta N =0$ is not suitable for the present framework because they do not satisfy the above-mentioned energy scale (the second assumption), we also present the numerical results for $\delta N = 0$ for completeness.
First, let us explain the overall structure of the phase diagrams within the single-$\bm{q}$ approximation (up to the zeroth order of the perturbation). 
There are two phase boundaries within the above calculation in each panel: one is the second-order phase boundary between the normal state and the superconducting state indicated by the blue line, and the other is the phase boundary between the two helical phases (or \red{between} the BCS state and the helical phase for $\delta N = 0$) indicated by the red line.
The latter boundary is first order, as indicated by the dashed lines, for $\delta N \ne 0$ and terminates inside the superconducting phase, while it is either first order (dashed line) or second order (solid line) for $\delta N = 0$.
The first-order transition line is determined by the comparison of the free energies $F_{\mathrm{SN}}$ between two helical phases, and on this line the Cooper pair momentum, $Q$, discontinuously changes.
We present the $\AM$ dependence of $Q$, at $T=0.1\Tc$ for different values of $\delta N$ by the solid lines in Fig.~\ref{fig:AMdep_of_momentum}.
For $\delta N=0$ case, $Q$ is zero in the small $\AM$ region,
and as $\AM$ increases, it exhibits the first-order transition to the helical phase with $Q\ne 0$. 
In this case, the inversion symmetry is not broken, and the helical phases with $Q$ and $-Q$ are degenerate [Fig.~\ref{fig:AMdep_of_momentum}(a)], and in the figure, we choose the positive $Q$. 
For $\delta N\ne 0$ case, $Q$ is not zero even in the small $\AM$ region except for $\AM=0$ because of the broken inversion symmetry [Figs.~\ref{fig:AMdep_of_momentum}(b) and \ref{fig:AMdep_of_momentum}(c)].
It also exhibits the first-order transition to larger $Q$ with increasing $\AM$ at $T=0.1 \Tc$. 
The first-order transition lines terminate at intermediate temperatures, above which the discontinuous changes become continuous (second-order for $\delta N = 0$ and crossover for $\delta N \neq 0$), as mentioned above.
\par
Next, we investigate the instability toward the stripe phase against the helical phase within the perturbative framework formulated in the previous section. 
\red{In this approach, we assume that the transition between the helical phase and the stripe phase is second order; if the transition is first order, the phase boundary cannot be determined within this framework.
In Sec.~\ref{conclusion}, we discuss the possibility that this assumption may break down.}
In Fig.~\ref{fig:phase_diagrams}, the parameters $(\AM, T)$ such that $\epsilon_1(\qst) = 0$ are plotted by the solid circles, and their colors represent the ratio $\ratio$.
We find the stripe phase in the parameter region characterized by the temperatures below and the altermagnetic splitting above the critical point. 
For $\delta N=0$ case [Fig.~\ref{fig:phase_diagrams}(a)] in a large $\AM$ region, both the helical phase and the stripe phase appear, as in Ref.~\cite{Agterberg-helicalstripe-2007}.
Considering that the Zeeman magnetic field favors the LO state rather than the FF state in the absence of the RSOC (i.e., in the presence of the inversion symmetry), the emergence of the helical phase seems implausible for $\delta N = 0$.
This is probably because our framework cannot be applicable for $\delta N = 0$, where the interband paring is the dominant channel rather than the intraband pairing.
Interestingly, the boundary of the stripe phase is nonmonotonic as a function of $\AM$ in the temperature range $0.2\lesssim T/\Tc\lesssim0.3$, which implies the reentrant behavior of the stripe phase.
For $\delta N\ne0$ [Figs.~\ref{fig:phase_diagrams}(b) and \ref{fig:phase_diagrams}(c)], the large portion of the phase diagram is occupied by the helical phase.
Compared with $\delta N=0$ case, the region of the stripe phase shrinks with increasing $\delta N$, reflecting the enhanced asymmetry between $\bm{q}$ and $-\bm{q}$.
This reduction is especially pronounced in the intermediate $\AM$ region $2.5\lesssim\AM/\Tc\lesssim3.0$.
This trend makes the reemergence of the stripe phase obvious in large $\delta N$ as shown in Fig.~\ref{fig:phase_diagrams}(c).
\par
Figure~\ref{fig:AMdep_of_momentum} also shows the minimizer of $\epsilon_1(\bm{q})$, \red{denoted by} $\bm{q}_{\mathrm{dip}}$, as a dashed line, and the difference $|\delta q| = ||Q| - |q_{\mathrm{dip}}||$ as a dash-dotted line.
The shaded region corresponds to the helical phase, where $\epsilon_1(\qdip)$ is positive, while the non-shaded region corresponds to the stripe phase, where it is negative.
Note that $q_{\mathrm{dip}}$ is equal to $q_{\mathrm{st}}$ at the phase boundary of the stripe phase, where $\epsilon_1(\bm{q}_{\mathrm{dip}})=0$.
Thus, in the following discussion, we approximately treat $q_{\mathrm{dip}}$ as a plausible momentum $q_{\mathrm{st}}$ for the stripe phase, even inside the phase boundary.
Our results indicate that $q_{\mathrm{st}}$ is present only in the $\AM$ region beyond the first-order transition point \red{between the two helical phases}.
While we find $|\delta q| \approx0.2$ just above the first-order transition, it goes to zero as $\AM$ increases, namely $q_{\mathrm{st}}\approx-Q$ in the large $\AM$ region.
\par
We discuss the superconducting gap structure of the stripe phase.
The absolute values of the ratio $\ratio$ shown in Fig.~\ref{fig:phase_diagrams} are calculated from the eigenstates of Eq.~\eqref{eq:linearized_gap_eq}, satisfying $\epsilon_1(\qst)=0$.
The ratio is at most 0.35 just above the first-order transition, and becomes negligibly small at large $\AM$ for all $\delta N$ values.
Thus, the higher harmonic component $\Delta_{2\bm{Q}-\qst}$ is smaller than $\Delta_{\qst}$ in the whole parameter region.
Considering $q_{\mathrm{st}}\approx-Q$ in the large $\AM$ region, it follows that the stripe phase is well described by $\Delta(\bm{R})\approx \Delta_{\bm{Q}}e^{i\bm{Q}\cdot\bm{R}}+\Delta_{-\bm{Q}}e^{-i\bm{Q}\cdot\bm{R}}$.
This stripe phase resembles those predicted in previous theoretical studies on Rashba--Zeeman superconductors~\cite{Aoyama_2024, Agterberg_Babaev_Garaud_2014}.
By contrast, the stripe phase in the small $\AM$ region exhibits relatively large values of $\ratio$ and $\delta q$.
Later, we will revisit the large-$\delta q$ behavior, which has not been observed even in Rashba--Zeeman superconductors.
\par
\begin{figure}[tb]
    \centering
    \includegraphics[width=0.9\linewidth]{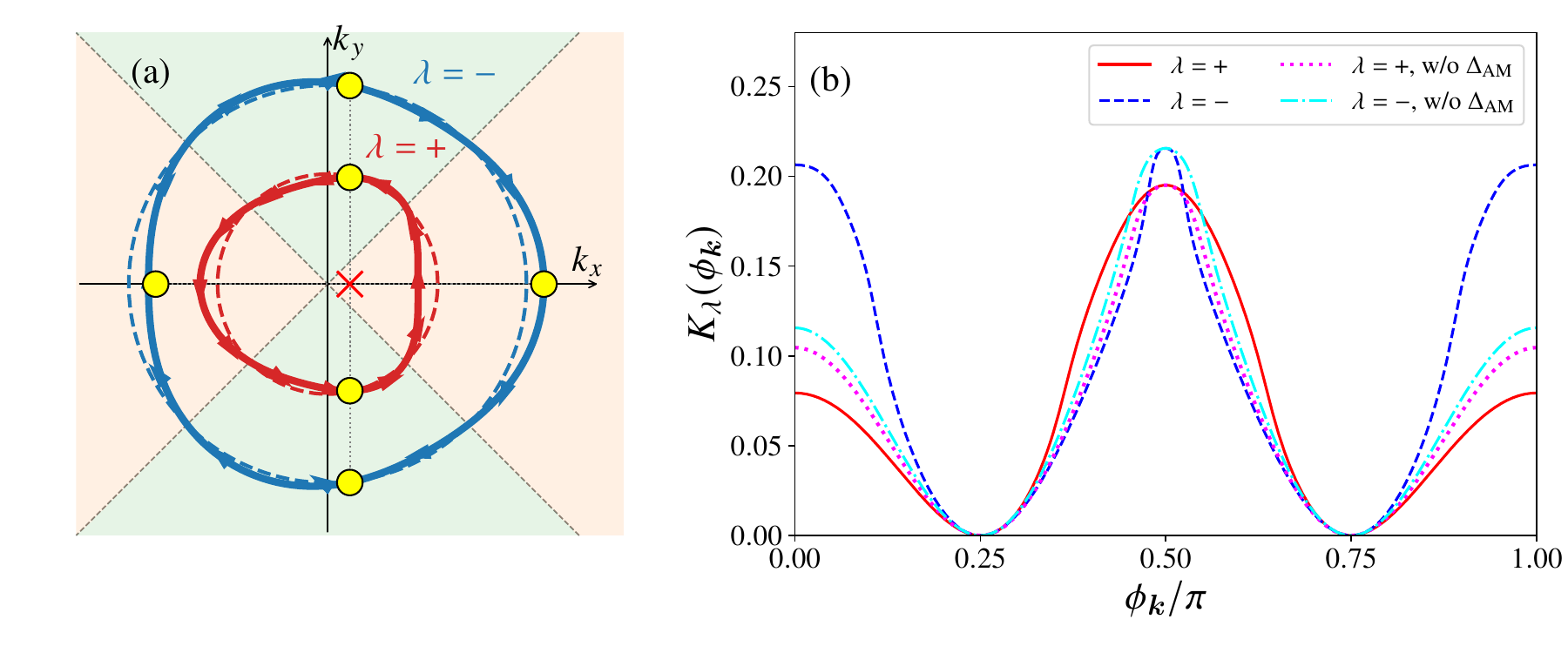}
    \caption{\red{(a) Schematic FSs of our model. The dashed lines represent the FSs with the RSOC term only \blue{($\AM=0$)}, while the solid lines show those with both the RSOC and altermagnetic splitting terms. Filled yellow circles indicate electrons forming Cooper pairs with the momentum denoted by the red cross mark. The dashed gray line shows the directions in which the altermagnetic splitting is absent. \blue{The orange (green) region corresponds to range A (B) introduced in the main text.}
    (b) $\phi_{\bm{k}}$ dependence of the integrand on the R.H.S. of Eq.~\eqref{eq:gapeq_in_helical} for $\delta N=0.05$, $T=0.1\Tc$ and $\AM=4.0\Tc$. The red solid (blue dashed) line shows the contribution from the inner (outer) FS, while the magenta dotted (cyan dash-dotted) line shows the corresponding inner (outer) FS contribution obtained by setting $\AM=0$ in $\omega_{\bm{Q},\lambda}$ appearing in  Eq.~\eqref{eq:gapeq_in_helical} with all other parameters unchanged.}}
    \label{fig:helical_pairing}
\end{figure}

Before investigating the stripe instability in detail, let us review the \red{dominant pairing channel} of the helical phase, in which both FSs ($\lambda=\pm$) cooperatively contribute to the pairing~\cite{Mukasa_Masaki_2025}.
As seen from Eq.~\eqref{eq:Normal_H0}, the RSOC and the altermagnetic splitting are coupled via the spin-$y$ component ($\sigma_y$). 
This coupling induces deformations of the FSs along the $k_x$ direction in the present system as shown in Fig.~\ref{fig:helical_pairing}(a).
Importantly, due to the anisotropy of the altermagnetic splitting, \red{both} FS\red{s (solid lines)} exhibit deformations depending on the momentum of electrons \red{from} the isotropic FS in the RSOC-only case \red{(dashed lines).} 
\blue{For convenience, we refer to the angular ranges $-\pi/4 < \phi_{\bm{k}} < \pi /4$ and $3\pi/4 < \phi_{\bm{k}} < 5\pi /4$ as range A [the orange region in Fig.~\ref{fig:helical_pairing}(a)], and the ranges $\pi/4 < \phi_{\bm{k}} < 3\pi /4$ and $5\pi/4 < \phi_{\bm{k}} < 7\pi /4$ as range B [the green region in Fig.~\ref{fig:helical_pairing}(a)].} 
\red{In the outer FS  (the solid blue line, $\lambda = -$)} in \blue{range A}, the deformation occurs toward the $+k_x$ direction, while in \blue{range B} it occurs toward the $-k_x$ direction.
The inner FS (the solid red line, $\lambda = +$) shows the opposite deformation pattern.
\red{To see how the FSs contribute to the helical phase under the above-mentioned deformation, Fig.~\ref{fig:helical_pairing}(b) shows $K_{\lambda}(\phi_{\bm{k}})$ in Eq.~\eqref{eq:gapeq_in_helical} as a function of $\phi_{\bm{k}}$ for $\delta N = 0.05$, $T = 0.1\Tc$, and $\AM = 4.0\Tc$.
The $\phi_{\bm{k}}$ dependence of $K_{\lambda}(\phi_{\bm{k}})$ identifies the dominant regions of the FS contributing to the formation of the helical phase.
Note that $K_{\lambda}(\phi_{\bm{k}})$ is symmetric with respect to $\phi_{\bm{k}}=\pi$, reflecting the mirror symmetry of the FSs with respect to the $k_x$ axis [see Fig.~\ref{fig:helical_pairing}(a)].
Figure~\ref{fig:helical_pairing}(b) also shows the contributions obtained by setting $\AM=0$ with all other parameters including $\bm{Q}$ and $\Delta_{\bm{Q}}^{(0)}$ unchanged in order to see how the deformation stabilizes the helical phase.}
\blue{In range A, $\AM$ noticeably modifies $K_{\lambda}(\phi_{\bm{k}})$ for both $\lambda = \pm$: the contribution is enhanced for the outer FS, while it is suppressed for the inner FS. In this angular range, the deformation of the outer (inner) FS occurs along (opposite to) $\bm{Q}$. In range B, a nonzero $\AM$ does not affect $K_{\lambda}(\phi_{\bm{k}})$ at $\phi_{\bm{k}} = \pi/2$ and $3\pi/2$, because $\hat{\bm{g}}(\bmkF)\cdot \bm{J}(\bmkF) = 0$ for $\bm{n} = \bm{e}_y$. Away from these angles, the contribution from the outer (inner) FS is slightly reduced (enhanced). 
In other words, as a consequence of the deformations parallel to $\bm{Q}$, the helical phase with $Q > 0$ is cooperatively stabilized by the outer FS in range A and the inner FS in range B, although the contribution from the outer FS in range B also exists independently of the deformation.
These dominant pairing channels are schematically sketched in Fig.~\ref{fig:helical_pairing}(a), in which the positions of electrons, indicated by the yellow circles, correspond to the peak angles in Fig.~\ref{fig:helical_pairing}(b).
}
\begin{figure}[tb]
    \centering
    \includegraphics[width=\linewidth]{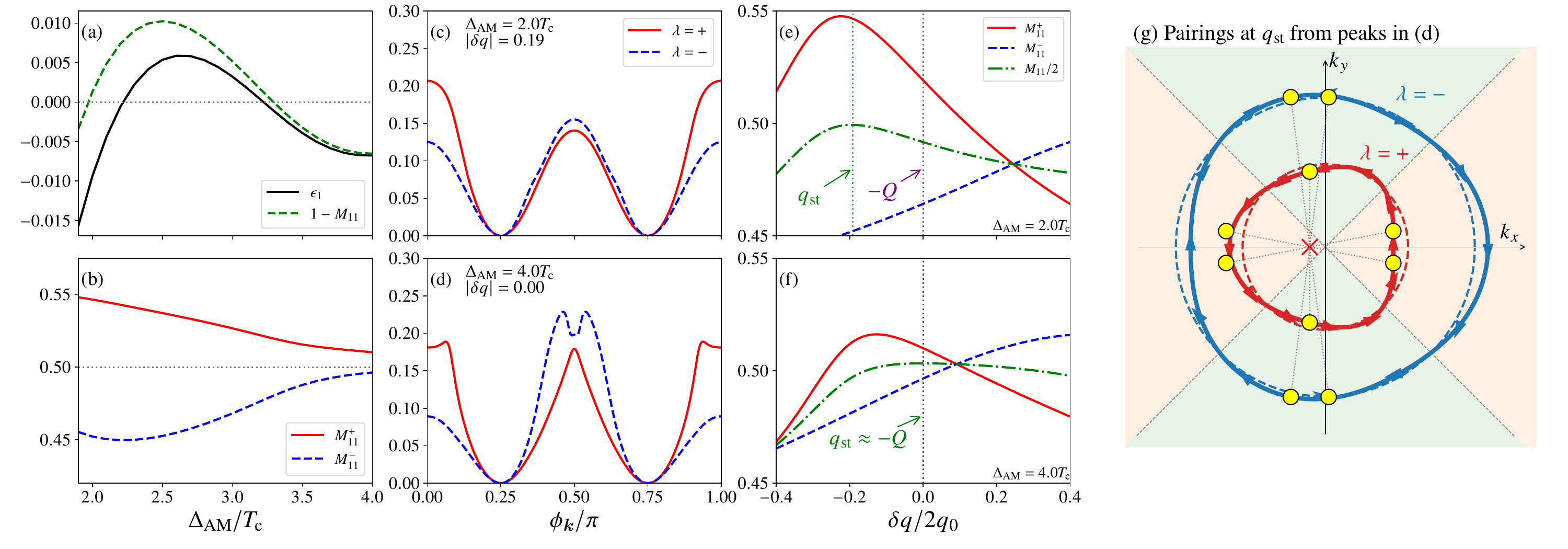}
    \caption{\red{(a, b) $\AM$ dependence of $\epsilon_1(\qst)$, $1 - M_{11}(\qst)$, and $M^{\pm}_{11}(\qst)$. $M^{\pm}_{11}$ is the contribution from $\lambda=\pm$ band and $M_{11}=M^{+}_{11}+M^{-}_{11}$. (c, d) $\phi_{\bm{k}}$ dependence of the integrand of $M^{\pm}_{11}$ for two $\AM$ values. (e, f) $\delta q$ dependence of $M_{11}^{\pm}$. The dotted green (purple) lines indicate the position of $q_{\mathrm{st}}$ ($-Q$). \red{We set $\delta N=0.05$ and $T=0.1\Tc$.} (g) Schematic illustration of the FSs with Cooper pairs having $q_{\mathrm{st}}$. The positions of the electrons are estimated from the peaks in (d). Definitions of lines are the same as those in Fig.~\ref{fig:helical_pairing}(a). }}
    \label{fig:schematic_FS}
\end{figure}
\par
Next, we describe the mathematical background underlying the instability toward the stripe phase. 
Focusing on the relative magnitudes of the matrix elements of $\mathsf{M}(\qst)$, we numerically find that $\epsilon_1(\qst)\approx 1-M_{11}(\qst)$ provides a good approximation especially in the large $\AM$ region, as shown in Fig.~\ref{fig:schematic_FS}(a).
To further clarify the origin of $M_{11}$, we decompose $M_{11}$ into contributions from each FS as $M_{11}=M^{+}_{11}+M^{-}_{11}$ based on the definition of $M_{11}$ in Eq.~\eqref{eq:def_M} and show them in Fig.~\ref{fig:schematic_FS}(b).
The solid red and dashed blue lines show contributions from the inner ($\lambda = +$) and outer ($\lambda = -$) FSs for $q_{\mathrm{st}}$, respectively.
While $M^{+}_{11}$ decreases monotonically, $M^{-}_{11}$ behaves nonmonotonically as a function of $\AM$.
This nonmonotonic behavior characterizes
$\epsilon_1$ in Fig.~\ref{fig:schematic_FS}(a), inducing its sign change, namely the reentrant structure of the stripe phase.
To elucidate this $\AM$ dependence of $M_{11}^{\pm}$, Figs.~\ref{fig:schematic_FS}(c) and \ref{fig:schematic_FS}(d) show the integrand of $M^{\pm}_{11}$ as a function of $\phi_{\bm{k}}$, after the summation over the Matsubara frequencies.
\red{Because of the mirror symmetry}, we show it only for $0\le\phi_{\bm{k}}\le\pi$.
We also show the momentum dependence of $M^{\pm}_{11}$ in Figs.~\ref{fig:schematic_FS}(e) and \ref{fig:schematic_FS}(f), where
$q_{\mathrm{st}}$ and $-Q$ are respectively indicated by the dotted green and purple line\red{s}.
\red{Figure~\ref{fig:schematic_FS}(g) schematically shows relation between the deformed FSs and the Cooper pairs with $q_{\mathrm{st}}$.
Positions of pairings are estimated by peaks in Fig.~\ref{fig:schematic_FS}(d).}
\par
\red{To discuss the difference between the small- and large-$\AM$ stripe phases that form the reentrant structure,} we analyze the stripe phase for $\AM=2\Tc$ and $\AM=4\Tc$ separately, and reveal each \red{dominant pairing channel} and how $q_{\mathrm{st}}$ is determined, based on the above results.
For $\AM=2\Tc$, the pairing with $q_{\mathrm{st}}$ is dominated by the electrons on the inner FS around $\phi_{\bm{k}}=0$ and $\pi$ [Fig.~\ref{fig:schematic_FS}(c)], where the FS deforms toward the $-k_x$ direction.
In this case, $M^{+}_{11}$ is sufficiently large compared with $M^{-}_{11}$ that the $\bm{q}$ dependence of $M_{11}$ is primarily governed by $M^{+}_{11}$ [Fig.~\ref{fig:schematic_FS}(e)].
As a result, $q_{\mathrm{st}}$ is optimized at the peak of $M^+_{11}$, which is deviated from $-Q$.
Next we examine the case of $\AM = 4\Tc$.
Notably, the pairing with $q_{\mathrm{st}}$ receives substantial contributions not only from the same region as in the $\AM = 2\Tc$ case, but also from the outer FS around $\phi_{\bm{k}}=\pi/2$ and $3\pi/2$ [Fig.~\ref{fig:schematic_FS}(d)], where the FS similarly shifts toward the $-k_x$ direction.
Since $M^+_{11}$ and $M^-_{11}$ become closer, the peak of $M_{11}$ is broadened and shifted toward $-Q$ [Fig.~\ref{fig:schematic_FS}(f)].
It is worth noting that the large contribution from the outer FS is indeed caused by the anisotropic deformation by the altermagnetic splitting.
In the Rashba--Zeeman superconductors, the outer (inner) FS shifts only toward $+k_x$ ($-k_x$) direction.
Thus, for example, the outer FS has small contributions to the pairing with the negative momentum.
\par
The reentrant behavior shown in Fig.~\ref{fig:phase_diagrams} can be attributed to the \red{$\AM$-dependent dominant pairing channel}s as observed in the above\red{-mentioned} two cases.
The most significant difference between them is whether the outer FS contributes substantially to the pairing with $q_{\mathrm{st}}$.
\red{In a small altermagnetic field the stripe phase is formed primarily in the inner FS.
As $\AM$ increases, the contribution from the outer FS is enhanced and finally in a large altermagnetic field the stripe phase is formed both in the inner and outer FS.}
In other words, the \red{dominant pairing channel} of the additional momentum is highly dependent on $\AM$, which is also accompanied by the change of the optimized $q_{\mathrm{st}}$.
This $\AM$ dependence of the \red{dominant pairing channel} characterizes the nonmonotonic behavior of $\epsilon_1$ and $M_{11}$ [Fig.~\ref{fig:schematic_FS}(a) and \ref{fig:schematic_FS}(b)], leading to the reentrant structure of the stripe phase.
\par
\section{Conclusion}\label{conclusion}
In this paper, we explore the possibility of the superconducting stripe phase in the altermagnet with the RSOC. 
Using the quasiclassical framework including multiple center-of-mass momenta of Cooper pairs, we numerically find the stripe phase in the low temperatures in the $(T,\AM)$ phase diagram.
Furthermore, the stripe phase shows the reentrant behavior as a function of $\AM$.
\red{We demonstrate that the stripe phase exhibits different properties in the small $\AM$ region, whcih is connected to the first-order phase boundary between the two helical phases, and in the large $\AM$ region, which is connected to the second-order phase boundary between the superconducting and normal phases.}
In the former parameter region, the additional momentum $\bm{q}_{\mathrm{st}}$ is deviated from $-\Qheli$, 
and the third component with center-of-mass momentum $2\bm{Q} - \qst$ appears.
However, in the latter region, $\qst$ is approximately $-\Qheli$, and the superconducting order parameter virtually consists of two momenta $\Qheli$ and $-\Qheli$.
The difference between the two parameter regions originates from the anisotropic deformation of the Fermi surfaces due to the altermagnetic splitting, resulting in the reentrant behavior.
\par
Our framework is insufficient to calculate the amplitude of the additional \red{components of} superconducting gap because it only includes the linearized gap equation in terms of the additional components.
Thus, how the superconducting gap evolves with increasing $\AM$ should be addressed within \red{an} advanced quasiclassical framework involving higher-order perturbative terms.
\red{We also remark on the possibility of a first-order phase transition between the stripe phase and the helical phase. In the phase diagrams shown in Fig.~\ref{fig:phase_diagrams}, most of the first-order phase transition lines (red dashed lines) may be replaced by a first-order phase transition line between the small-$\AM$ helical phase and the stripe phase, located below the original red dashed lines.
Correspondingly, the assumption that the phase transition is second order may break down and the second-order phase boundary between the stripe phase and the helical phase, indicated by the instability line, may partly turn into a first-order phase transition in the vicinity of the red dashed lines. 
Because the present perturbative approach cannot describe first-order phase transitions to or from the stripe phase, these issues  are  beyond the scope of the present study and require the advanced quasiclassical framework.}
It is also necessary to fully calculate the real-space structure of the finite-momentum superconducting phases suggested in this work in order to provide evidence that can be experimentally detected. 
\backmatter
\bmhead{Acknowledgments}
K. M. acknowledges financial support by a research granted from Murata Science and Education Foundation.
This work was also supported by JSPS KAKENHI Grant Numbers JP24K17000 and JP23K22492.








\bibliography{sn-bibliography}

\end{document}